\documentstyle[12pt,epsfig]{article}
\begin{document}

\title{Using Immunology Principles for Anomaly Detection in Electrical Systems}
\author{P. J. Costa Branco, J. A. Dente and R.Vilela Mendes\medskip \thanks{%
e-mails: pbranco@alfa.ist.utl.pt, edentepc@alfa.ist.utl.pt,
vilela@cii.fc.ul.pt} \and {\small Laborat\'{o}rio de Mecatr\'{o}nica, DEEC,
Instituto Superior T\'{e}cnico,} \and {\small Av. Rovisco Pais, 1096 Lisboa
Codex, Portugal} \and {\small Zentrum f\"{u}r interdisziplin\"{a}re
Forschung, Universit\"{a}t Bielefeld,} \and {\small Wellenberg 1, 33615
Bielefeld}}
\date{}
\maketitle

\begin{abstract}
The immune system is a cognitive system of complexity comparable to the
brain and its computational algorithms suggest new solutions to engineering
problems or new ways of looking at these problems. Using immunological
principles, a two (or three-) module algorithm is developed which is capable
of launching a specific response to an anomalous situation. Applications are
being developed for electromechanical drives and network power transformers.
Experimental results illustrate an application to fault detection in
squirrel-cage electric motors.
\end{abstract}

\section{Introduction}

The immune system, with its cell diversity and variety of information
processing mechanisms, is a cognitive system of complexity comparable to the
brain. Understanding the way this organ solves its computational tasks
suggests new engineering solutions or new ways to look at old problems.

Early fault detection and predictive maintenance are extremely important for
the cost savings they provide, especially in large and complex systems with
many pieces of equipment, as an electrical power network. In a system of
such complexity (because of its many connections and diversity of equipment)
it is difficult to make a complete catalog of all the possible, and
probable, anomalous situations. On the other hand when, for example, a
circuit breaker operates in response to an overload, the harm is already
done, either to some equipment or to the stability of the network. With its
ability to detect and react to novel situations and to unleash smooth early
secondary responses, the immune system seems to be an adequate source of
inspiration to develop algorithms for early detection of anomalous behavior
in electrical systems.

After a short introduction to the characteristics of the immune system that
are relevant for our purposes, the paper describes a three-module algorithm
for anomaly detection. By analogy with the immune system, these modules will
be called the $B$-module, the $T$-module and the $D$-module. However, it
should be emphasized that we are not trying to imitate the immune system in
all its features and detailed operation. Many of the detailed features of
the immune system are dependent on the biological context where it operates
and on the type of cell hardware that is used. It is our opinion that the
correct way to profit from the clever evolutionary mechanisms developed by
Nature is to obtain algorithmic inspiration from them but, at the same time,
to find the implementation that is more appropriate for our problem. For
example, the interaction between the $B$-module and the $T$ -module takes
the reverse order of what is found in Nature, with a clone proliferation
phase preceding $T$-identification.

The algorithmic modules developed in Section 2 have a wide range of
applications to many different technological systems. As an illustration, a
concrete case is dealt with in the last part of the paper, namely an
application to fault detection in squirrel-cage electric motors. Another
application that is being currently developed concerns aging and anomaly
detection in power transformers of an electrical network.

\section{The immune system paradigm}

Some of the immune system features that are of interest for anomaly
detection in complex technological systems are: \cite{Complex} - \cite
{Timmis}

\begin{itemize}
\item  {\it Uniqueness}: the immune system of each individual is unique,
each one being a different entity, in spite of their overall similarity.
Similar pieces of equipment are also unique entities. For instance, electric
motors of same type and with equal ratings have different aging processes
when placed in different electrical and thermal stress conditions. Therefore
they also require a protection system that is tuned to their specific
vulnerabilities.

\item  {\it Imperfect detection and mutation}: by not requiring an
absolutely precise identification of every pathogen, the immune system
becomes flexible and increases its detection range. However, when a pathogen
is detected, an hypermutation mechanism sharpens the identification. Because
identification of pathogens is made by partial matching, a small number of
the ``detectors'' ($10^{8}$ to $10^{12}$) is able to recognize non-self
patterns on the order of $10^{16}$. Similarly, it is not an easy task to
characterize precisely all the possible anomaly situations in a complex
system. Therefore, an initial rough characterization of the anomalies and
imperfect detection seems an useful feature. That is, initially a small
number of detectors may be defined, which are at a later stage modified by
the dynamics.

\item  {\it Learning and memory}: the immune system is able to learn the
structure of the pathogens, and remember those structures. Future responses
are much faster and, when made at an early stage of the infection, no
adverse effects are felt by the organism. We underline the importance of
this feature for smooth operation and cost savings, both in fault detection
and in preventive maintenance.

\item  {\it Novelty detection}: the immune system can detect and react to
pathogens that the body has never encountered before.

\item  {\it Distributed detection}: the detectors used by the immune system
are small and efficient, highly distributed and not subjected to centralized
control.
\end{itemize}

\section{The algorithm}

In the algorithm, the states of the system, both {\it normal condition} and 
{\it anomaly} states, are characterized by the values of $N$ variables. The $%
N-$dimensional state vector is normalized in such a way that all variables
take values in the interval $[0,1]$. The values of the state vector in
normal conditions define the {\it self }$S$ of the system. The anomaly
states are the {\it non-self} of the system.

The algorithm contains three modules. The $T$-module discriminates self from
non-self (that is, from anomalies). The $B$-module reacts to all frequently
occurring state vector values (self and non-self codes) and presents its
results to the $T$-module. The $B$-module also plays a role in updating the $%
T$-module. For large distributed systems one considers also the
implementation of $D$-modules which are essentially reduced state space $T$%
-modules. They are simple anomaly detection units that, when some potential
abnormal situation is detected, report the situation to a central system
which then makes a detailed analysis of the event.

\begin{figure}[htb]
\begin{center}
\psfig{figure=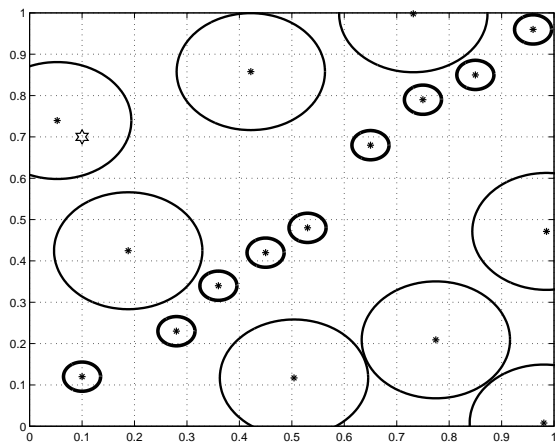,width=9truecm}
\end{center}
\caption{The $T$-module structure: Self patterns (small circles) and anomaly
detectors (large circles)}
\end{figure}

\subsection{The T-module}

This module contains a set of detectors which are vectors in complementary
space of the self, $A=[0,1]^{N}\backslash S$. Each element $\overrightarrow{x%
}$ of $A$ is able to detect anomalies inside a radius $r_{x}$ around it.
That is, if $|\overrightarrow{y}-\overrightarrow{x}|<r_{x}$, $%
\overrightarrow{y}$ being the current state of the system, an anomaly of
type $x$ is reported.

Let the set of self patterns be known. Then, the algorithm defines $d$
detectors with radius $r_{x}$. Fig.1 illustrates, in a two-dimensional
space, the basic idea. The small circles are the self patterns. To each
point in the self corresponds a code (a set of vector coordinates) and an
affinity neighborhood of normal operating conditions inside a radius $r_{s}$%
. The anomaly detectors are shown in the figure as large circles. The $T$%
-module is initialized by choosing points in $A$ at random with
corresponding radius $r_{x}$, until a reasonable coverage of the space $A$
is obtained with $d$ detectors.

The $T$-module receives inspiration from the censoring mechanism for {\it T}%
-cells that occurs in the thymus. This negative selection process is
implemented in the $T$-module as follows: Each candidate anomaly detector
(immature $T$-detector) is generated at random. The affinity of this vector
with those defining the self and the other already established detectors is
measured by the Euclidean distance. If the new detector falls in the
neighborhood domain of another detector or of a self code, it is deleted and
another candidate detector is generated at random. Otherwise, the detector
is included as an element $\overrightarrow{x}$ of the space $A$ (mature $T$%
-detector). The censoring mechanism is repeated until a pre-specified number 
$d$ of detectors is generated.

When a measurement $\overrightarrow{y}$ of the system arrives to the $T$%
-module, the algorithm verifies whether this code has affinity with one of
the detectors or with the self. If it falls in the self domain, no detector
is activated. If affinity is found with one of the detectors $(%
\overrightarrow{x^{^{\prime }}})$, an anomaly of type $\overrightarrow{%
x^{^{\prime }}}$ is reported. To avoid over-reaction to spurious situations,
each detector is equipped with a counter and reports an anomaly only if the
offending vector occurs more than a predefined number of times.

We notice at this point that, with random initialization, the $T$-module is
not necessarily tuned to the most frequent, or probable, anomalous
situations. On the other hand, some dangerous gaps may occur between the
neighborhood domains of the detectors. This situation is corrected by
dynamic interaction with the $B$-module.

By a mechanism to be described later on, the $B$-module generates vector
codes corresponding to the most frequently occurring states of the system
and sends these codes as {\it alert codes} to the $T$-module. When an alert
code $\overrightarrow{\hat{x}}$\ coming from the $B$-module arrives to the $T
$-module, the latter takes one of three actions:

a) If $\overrightarrow{\hat{x}}$ is located inside a detector, the center of
this detector is shifted to a position closer to the $\overrightarrow{\hat{x}%
}$ code and, if $\overrightarrow{\hat{x}}$ continues to occur, an anomaly is
reported. That is, not only is an anomaly detected but also the detector
becomes better tuned to this kind of anomaly.

b) If $\overrightarrow{\hat{x}}$ is located outside all detector
neighborhoods, at a distance at least $r^{^{\prime }}$, a new detector is
created at the position of the $\overrightarrow{\hat{x}}$ code with radius $%
r^{^{\prime }}$. As before, if this situation recurs, an anomaly is reported.

c) Finally, if $\overrightarrow{\hat{x}}$ has affinity with a self pattern,
nothing happens.

\begin{figure}[htb]
\begin{center}
\psfig{figure=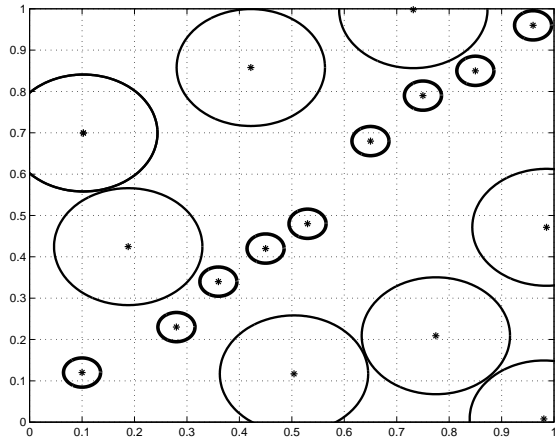,width=9truecm}
\end{center}
\caption{Shift of a detector code to increase affinity with an anomaly}
\end{figure}

\begin{figure}[htb]
\begin{center}
\psfig{figure=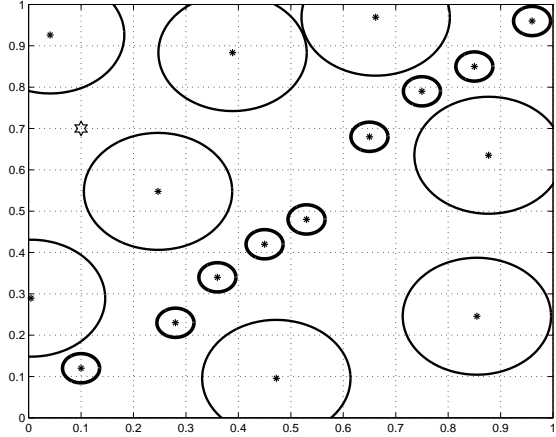,width=9truecm}
\end{center}
\caption{An anomaly outside all detector neighborhoods}
\end{figure}

\begin{figure}[htb]
\begin{center}
\psfig{figure=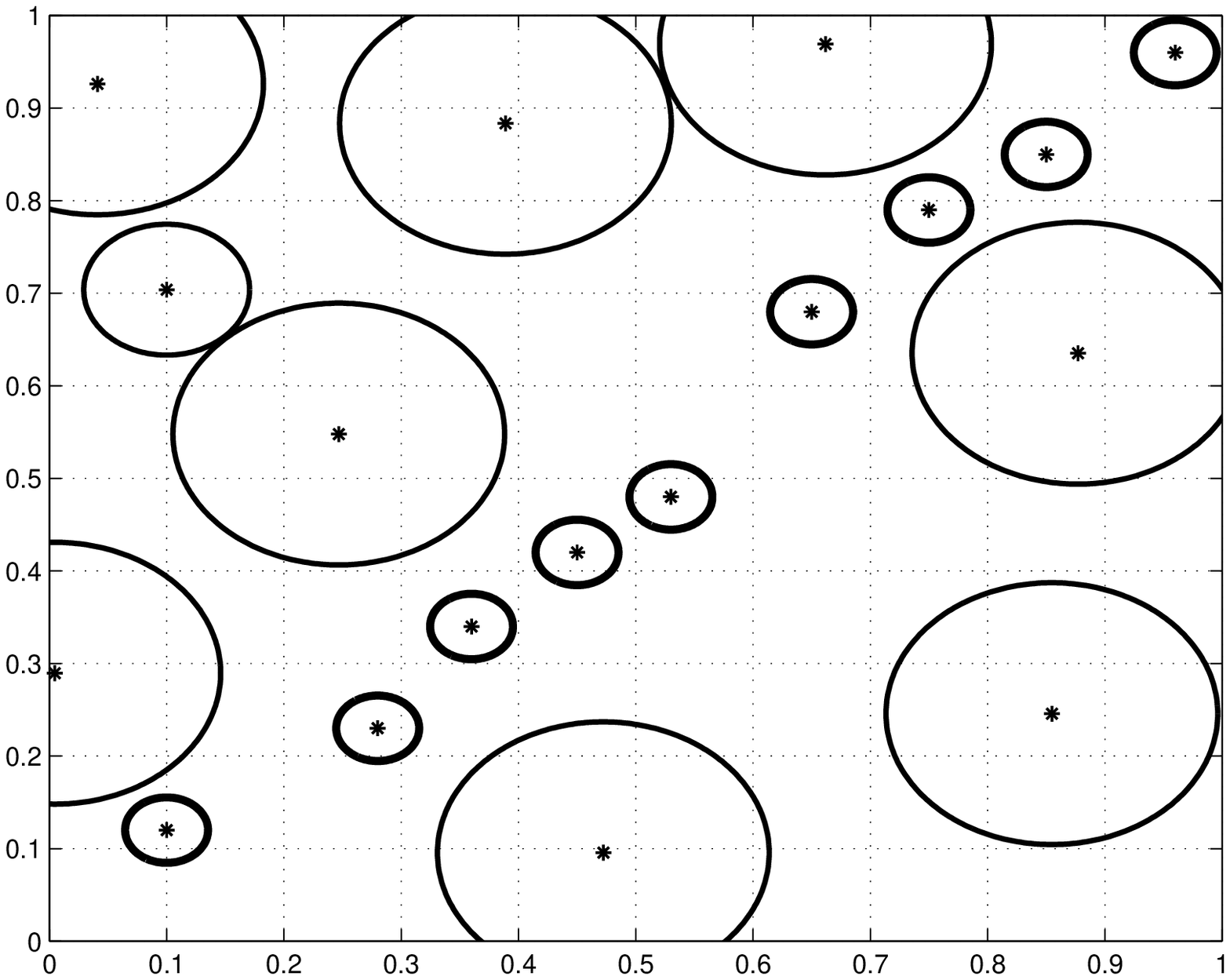,width=9truecm}
\end{center}
\caption{Creation of a new detector}
\end{figure}

By itself or in interaction with the $B$-module, the $T$-module is an
adaptive system. As an illustration, two typical situations are considered
here. The first one is shown in Figs.1 and 2. Suppose that a non-self code
(the star symbol in Fig.1) is detected several times. Then, the detector
changes its code to increase the affinity to this type of anomaly, as shown
in Fig.2. The second situation, shown in Fig.3, corresponds to a case in
which no detector had affinity with the external code. In this case, if the
situation occurs many times, the algorithm creates a new detector with a
resolution defined by the smallest distance to the other detector
boundaries, as shown in Fig.4. In this way the $T$-module modifies the
initial set of detectors produced by the censoring mechanism. It may change
their number, modify the space distribution and change the resolution,
creating a system-specific anomaly detection device.

\subsection{The B-module}

The $B$-module plays a role in improving the $A$ space coverage of the $T$%
-module and, when used with a system that is known to be operating in normal
conditions, may also be used to generate the self patterns.

The $B$-module has a total population of $N_{t}$ vectors 
\[
N_{t}=N_{l}+N_{l_{c}} 
\]
consisting of an initial population of $N_{l}$\ vectors $\left\{ 
\overrightarrow{x_{l}}\right\} $,\ randomly distributed in the whole space $%
[0,1]^{N}$, plus a variable number $N_{l_{c}}$ of clone vectors $\left\{ 
\overrightarrow{x_{l_{c}}}\right\} $. The number of clone vectors changes as
the system evolves. To keep the system computationally efficient their
number is limited to{\em \ }a fraction of the initial population 
\[
N_{l_{c\_\max }}=\beta N_{l} 
\]

The evolution of the total population is controlled by interaction with the
state vectors generated by the system. This dynamical evolution has {\it %
mutation} and {\it stimulation} features.

\subsubsection{Mutation}

A mutation process takes place every time an external code $\overrightarrow{y%
}$, coming from the system, arrives to the $B$-module. The mutation process
begins by selecting, from the total population, a sample of $p_{m}$ vectors, 
$\left\{ \overrightarrow{x_{m}}\right\} $. The mutation process operates
only in this part of the population and in those codes that are close to the
external signal $\overrightarrow{y}$.

\begin{figure}[htb]
\begin{center}
\psfig{figure=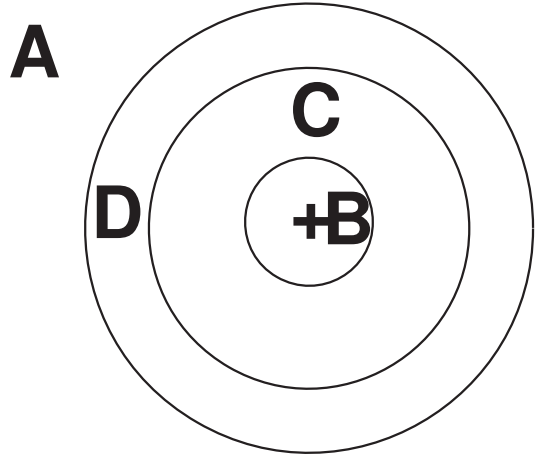,width=7truecm}
\end{center}
\caption{Zones {\it A} to {\it D} for the mutation process}
\end{figure}

The mutation process depends on the affinity between the vectors $%
\overrightarrow{x_{m}}$ in the sample and the external code $\overrightarrow{%
y}$. When the vector $\overrightarrow{x_{m}}$ and the code $\overrightarrow{y%
}$ are far way, as in zone {\it A} of Fig.5, no affinity is considered to
exist and the code $\overrightarrow{x_{m}}$ is not changed. Also, in zone 
{\it B,} there is no modification. For codes $\overrightarrow{x_{C}}$ in
zone {\it C,} the mutation process occurs in a deterministic way. The
external code $\overrightarrow{y}$ is assumed to have mass one and the
vectors in zone {\it C} mass $m_{l}$. The new code in the population
corresponds to the center of mass of the two entities, given by 
\[
\overrightarrow{x_{C}}\left( t+1\right) =\frac{m_{l}\overrightarrow{x_{C}}%
\left( t\right) +\overrightarrow{y}}{1+m_{l}} 
\]

In zone {\it D}, the mutation process works in a random way. The new
position of the population vector $\overrightarrow{x_{D}}$ is found using a
random uniform distribution for each point of the line defined by the old
position of the vector and the position of the external code 
\[
\overrightarrow{x_{D}}\left( t+1\right) =\overrightarrow{x_{D}}\left(
t\right) +\eta \left( \overrightarrow{y}-\overrightarrow{x_{D}}\left(
t\right) \right) 
\]
When the external code appears repeatedly in the same region, the mutation
process leads to a population cluster in that region. The speed of cluster
formation depends on the parameters $m_{l}$ and $\eta $.{\em \ }The cluster
code is computed by a hierarchical binary tree routine, a cluster being
identified when a threshold parameter in the clustering algorithm is
reached. The cluster center defines an {\it alert code} that is passed to
the $T$-module to be processed as described before. If the code is
identified as a self code it is eliminated from the population. Otherwise,
it reports an anomaly and creates a new or improved detector in the $T$%
-module.

\begin{figure}[htb]
\begin{center}
\psfig{figure=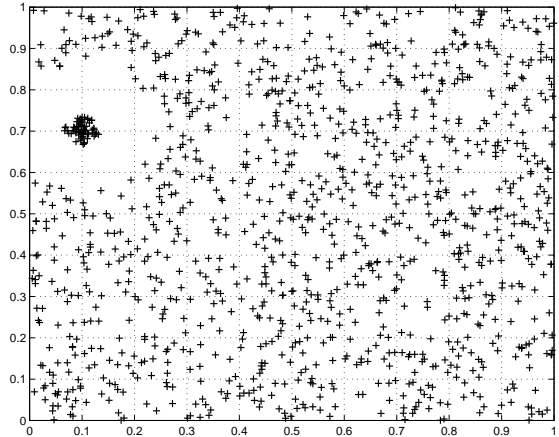,width=9truecm}
\end{center}
\caption{Mutation creates a cluster around an external code}
\end{figure}

The mutation process is illustrated in Fig.6 which corresponds to an
external code $\overrightarrow{y}=\left( 0.1,0.7\right) $ being detected
many times. A cluster is created around the external code.

It is also clear that, when either the nature of the self codes is not known
or it is difficult to specify, the $B$-module, used with a system in normal
operating conditions, may be used to generate the self codes.

\begin{figure}[htb]
\begin{center}
\psfig{figure=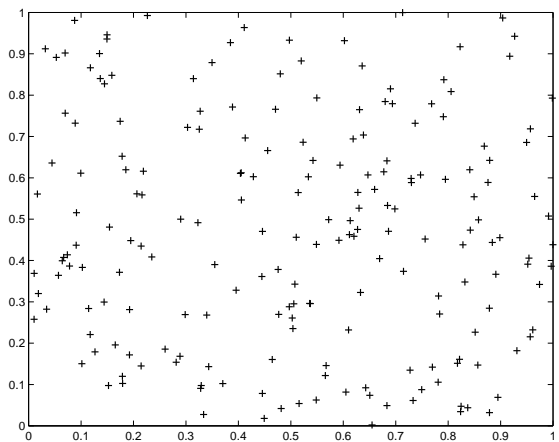,width=9truecm}
\end{center}
\caption{Initial vector population}
\end{figure}
\begin{figure}[htb]
\begin{center}
\psfig{figure=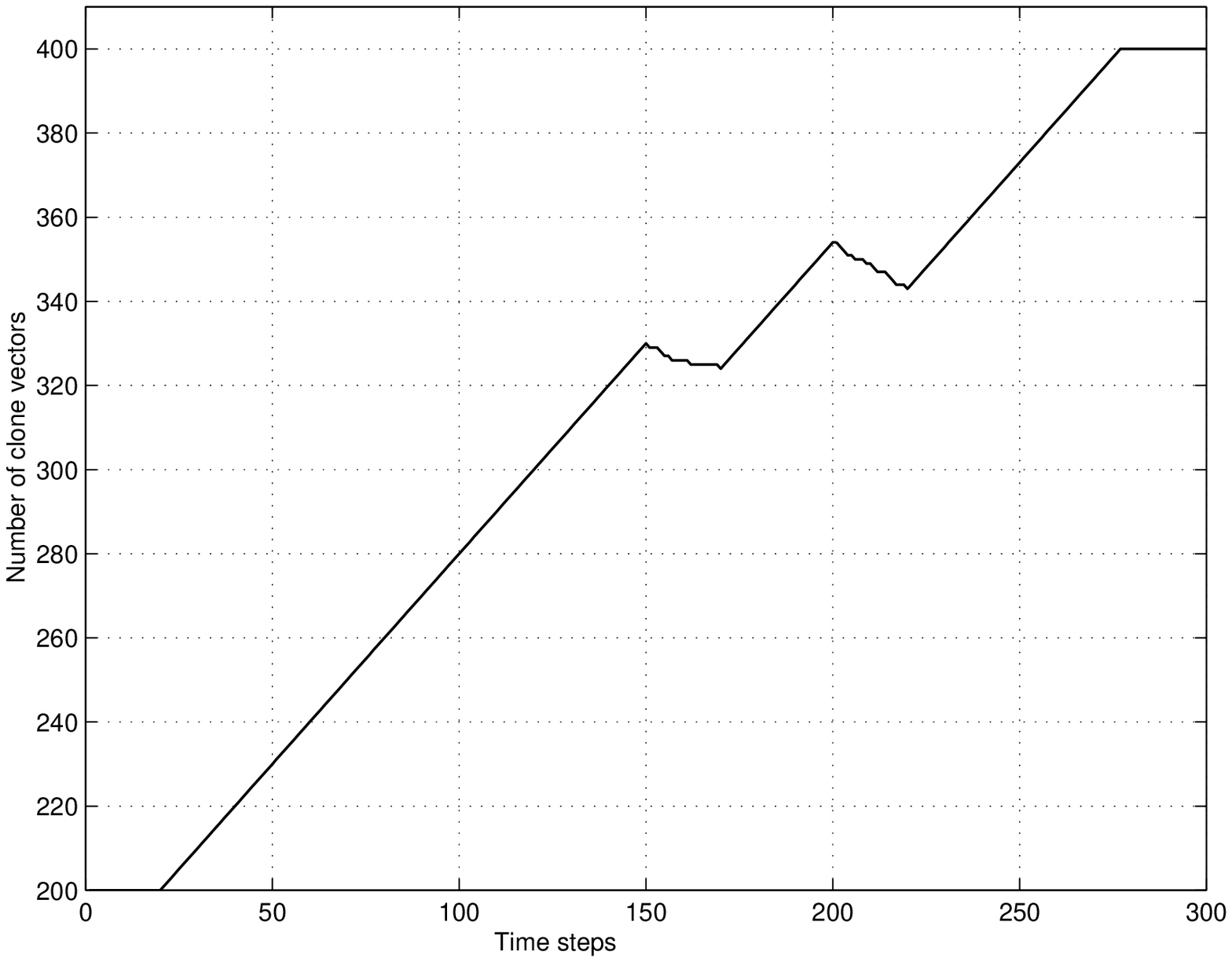,width=9truecm}
\end{center}
\caption{Evolution of the clone population}
\end{figure}

\subsubsection{Stimulation}

As new external codes arrive, the mutation process destroys the initial
uniformity of the population, resulting sometimes in a highly sparse
distribution. In this case, if an atypical external code appears, its
detection may become difficult. This situation is even more critical when
the external code is located near the boundary of the state-space or near an
already formed cluster. In these cases, the mutation process has a high
probability to fail because the areas {\it C} and {\it D} (Fig.5) are
depleted. To overcome this drawback, a {\it stimulation} or {\it cloning}
mechanism has been included in the algorithm to create new vectors in the
region where the external code appears.

\begin{figure}[htb]
\begin{center}
\psfig{figure=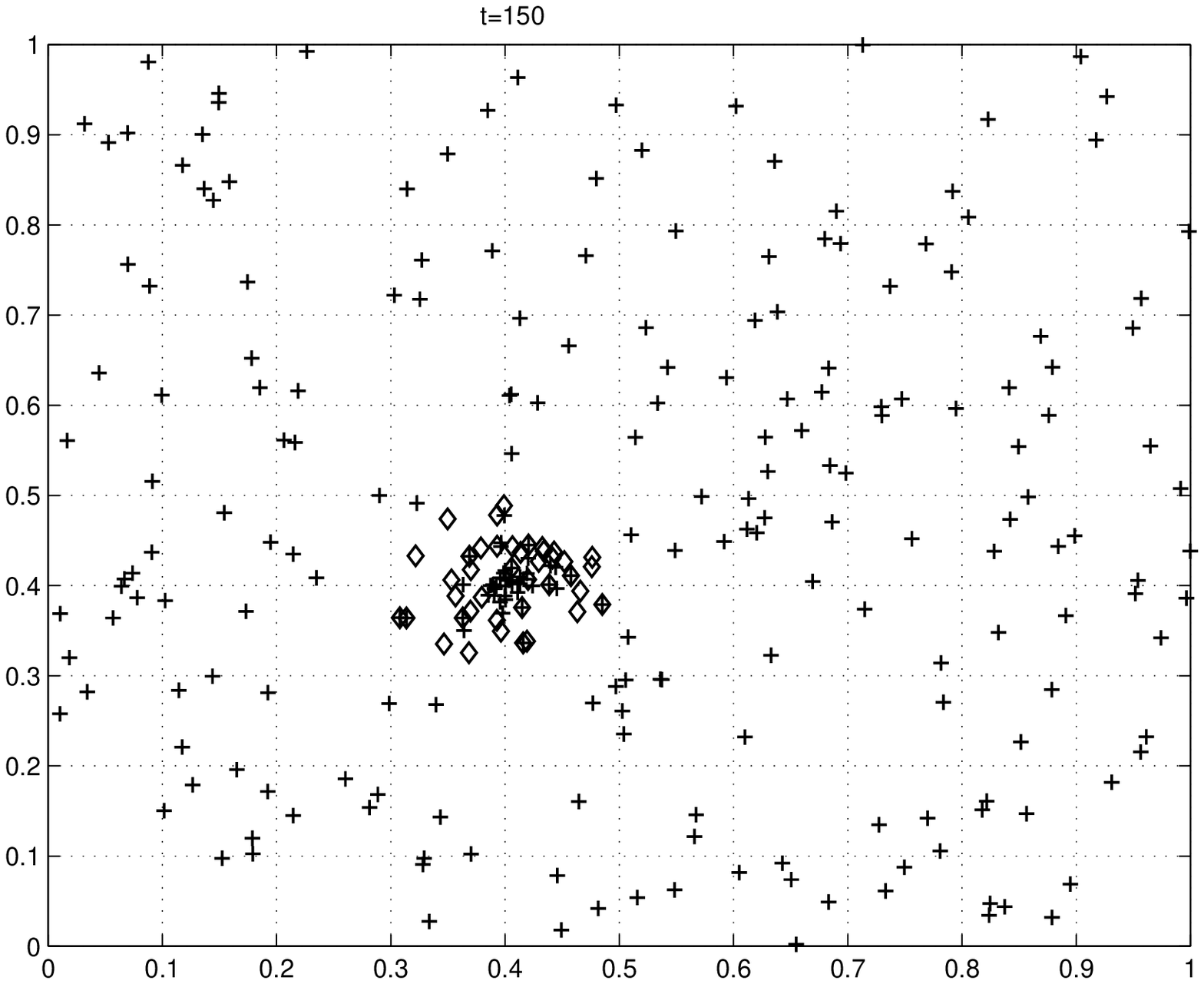,width=9truecm}
\end{center}
\caption{Population (initial $+$ plus clones $\Diamond $) at t=150}
\end{figure}
\begin{figure}[htb]
\begin{center}
\psfig{figure=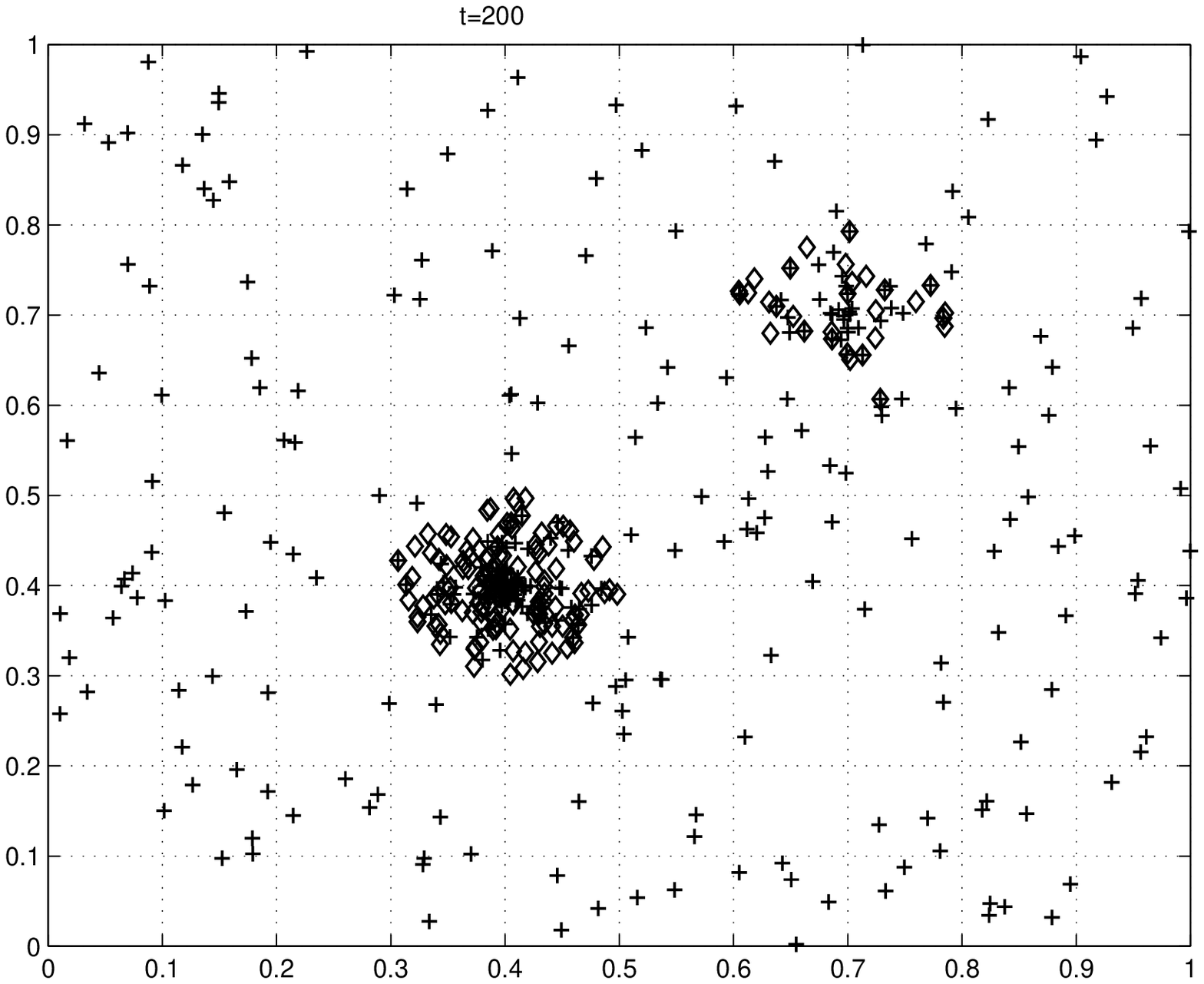,width=9truecm}
\end{center}
\caption{Same as Fig.9 at t=200}
\end{figure}
\begin{figure}[htb]
\begin{center}
\psfig{figure=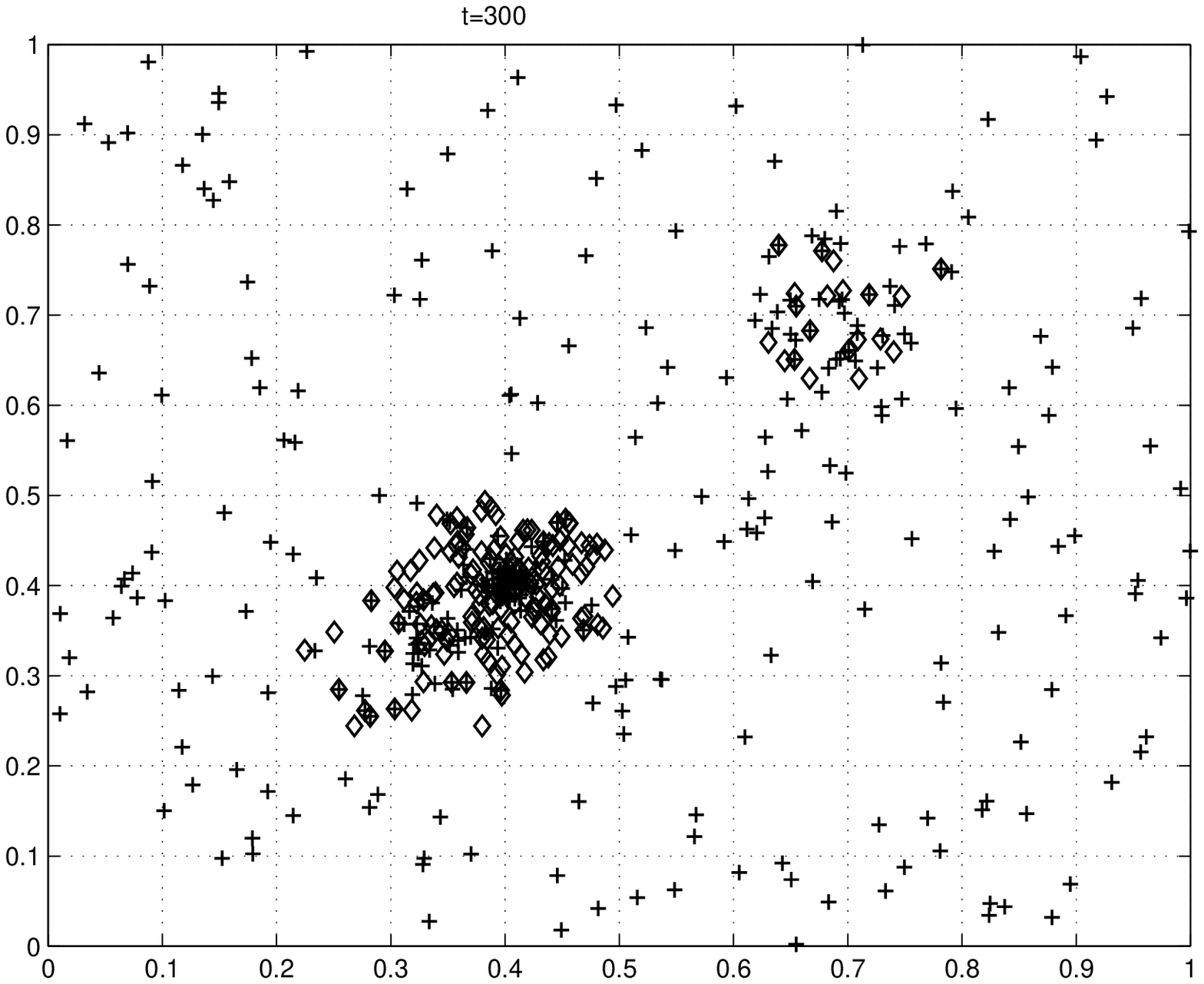,width=9truecm}
\end{center}
\caption{Same as Fig.9 at t=300}
\end{figure}

The cloning mechanism is activated when the rate of external codes arriving
in a neighborhood exceeds a pre-specified threshold $\alpha $. If the rate
is below this threshold there is no cloning and only the mutation process
takes place. The number of clone vectors in the total population is limited
to a maximum $N_{l_{c\_\max }}$. As long as the stimulation threshold is
exceeded in a given region, clones are added to the population in locations
chosen at random in the regions B to D. If the number of clone vectors has
reached its maximum value\ and more external codes continue arriving, at a
rate above the stimulation threshold, the oldest clone in the population is
replaced by the external code. After each cloning the mutation mechanism is
activated as before.

A death mechanism is also introduced for the clone vectors. At each time one
vector from the total population is chosen at random. If it belongs to the
initial $N_{l}$\ population nothing happens. However, if it belongs to the
set of $N_{l_{c}}$\ clone vectors, it is eliminated with probability 
\[
P_{die}=\frac{N_{l_{c}}}{N_{l_{c}}+N_{l}} 
\]

The stimulation mechanism is illustrated in Figs.7-11. The initial
population in Fig.7 was exposed from time $t=0$ to $t=150$ to the code $%
a=\left( 0.4,0.4\right) $, from $t=151$ to $t=200$ to the code $b=\left(
0.7,0.7\right) $ and from $t=201$ to $t=300$ to the code $c=\left(
0.32,0.32\right) $. Fig.8 shows the evolution of the total number of clone
vectors. Notice that clones only start to be created after a certain number
of occurrences of the external code. During a time interval, after the
change of external code, only the random death mechanism is observed. In
this example the saturation level $N_{l_{c}}$ is set at 400. Figs.9 -11 show
three successive snapshots of the total population. Clones are denoted by a
diamond symbol.

Memory is a hallmark of the immune system. As in the immune system, memory
features are present in the algorithm at two levels. In the $T$-module the
tuning of the position and range of the detectors acts as a long time memory
of past anomaly situations. For the $B$-module the distribution of the
population, both initial population and clones, has memory features at two
different time scales.

\subsection{D-modules}

For large distributed systems like, for example, an electrical power
distribution network it is not economic, in terms of computational power and
communications, to keep a permanent monitoring of the whole system by a
centralized facility. Therefore in these cases it is reasonable to
distribute throughout the network a large number of simple monitoring
systems, which report to the central unit only when some unusual condition
appears as a candidate anomaly. Depending on the nature of the transferred
data to the central unit, this one may then subject this particular network
node to a finer analysis. The $D$-modules behave therefore as {\it %
anomaly-presenting} systems and may be constructed as simplified $T$-modules
with a reduced state-space. For example, to monitor the power transformers
in a network, the $D$-modules may only process the data related to the
concentration of a particular gas type or the temperature at a few critical
points. In case of a potentially anomalous situation, the central unit may
then require the transfer of data from other monitoring devices. Although
simplified, the $D$-modules may also be evolving units, learning both from
their experience of local conditions and from periodic updates to benefit
from experience collected at other points of the network.

\section{Anomaly detection in squirrel-cage motors.}

Here, the algorithm is applied to fault detection in squirrel-cage motors.
These motors are critical components of many industrial equipments. They are
fairly reliable machines, nevertheless they do suffer degradation and
occasional failure. Environment, duty cycles and installation conditions may
combine to accelerate motor failure. Their malfunction introduces costs in
the industrial processes where they are inserted, these costs being often
much higher than the actual cost of the motor.

A variety of fault detection techniques have been proposed \cite{Penman} - 
\cite{Milimon}. Each one addresses a specific failure in one of the three
motor components: the stator, the rotor, or the bearings. The signals to be
monitored may be stator voltages and currents, output torque, rotor position
and speed, air-gap flux density, temperature and vibrations. A large amount
of research has concerned the stator current spectrum. The presence of some
frequency components has been shown to be the signature of a fault condition.

Most methodologies based on the stator current spectrum are based on the
assumption that the current drawn by a normal motor has only a significantly
large component at the supply frequency. A machine malfunction would
manifest itself by the appearance of other significant components, some
frequencies being related to specific faults. However, it is difficult to
distinguish in this way a normal operating condition from a potential
failure. This is because spectral components may arise from a number of
sources, including those related to normal operating conditions. Harmonic
components may exist, caused by the motor design, by the power network or by
fluctuations in the load torque, which are not related with an abnormal
motor condition.

To consider a single component spectra as the representative pattern of
normal motor condition is not correct. It is the actual current spectra,
obtained when the motor is inserted in the production unit in typical
conditions, that characterizes the {\it normal operating conditions}. It is
this frequency spectrum, with or without harmonics, that characterizes the 
{\it self} state of the machine.

\subsection{Fault detection: results and discussion}

The single-phase stator current is sampled and converted to the frequency
domain using a discrete Fourier transform. A sequence of spectra is obtained
by a moving window on the current data. Only those harmonics with amplitude
greater than $50$ dB and frequencies below $200$ Hz are kept. The $200$ Hz
upper limit was chosen because most frequencies related to squirrel-cage
motor faults occur in this interval. Harmonic amplitudes are normalized by
the amplitude of the supply frequency.

A detailed anomaly detection system must consider the overall frequency
spectrum pattern. Even discretizing the spectrum, this implies monitoring a
very high-dimensional space, a process that requires considerable computer
power. On the other hand, it is not necessary to be monitoring all
frequencies in full detail all the time. The solution is to segment our
anomaly-detection system into several subunits each one covering a frequency
interval in the spectrum. The frequency resolution of all the subunits need
not be same because, from experience, we already know what frequency
intervals need to be monitored in more detail. All the subunits work in a
similar fashion, the overall operating mode of the system being such that at
each time only one subunit is active but the active one changes
periodically. In this way an adequate fine covering of the energy spectrum
is implemented, while keeping the computation requirements at a reasonable
level.

A segmentation scheme of this type is applicable to any system where a good
resolution of the operation conditions is desired. However if, at the same
time, an overall continuous monitoring of the whole system is necessary, the
outputs of the subunits may be looked at as components of a global coding vector
that is sent for analysis to a central system. The central system may then
operate on a slower time schedule.

\subsubsection{Results}

The operation of the subunits is now illustrated using laboratory results
from normal conditions and from two typical fault situations.

Each subunit monitors a frequency interval around a frequency value $f_{D}$.
The interval is divided in two equal parts. For each part, the integral of
the spectrum is computed, obtaining a data pair that is sent as the external
code to the algorithm. The lower and higher intervals correspond to the absciss
and ordinate coordinates in the figures.

\begin{figure}[htb]
\begin{center}
\psfig{figure=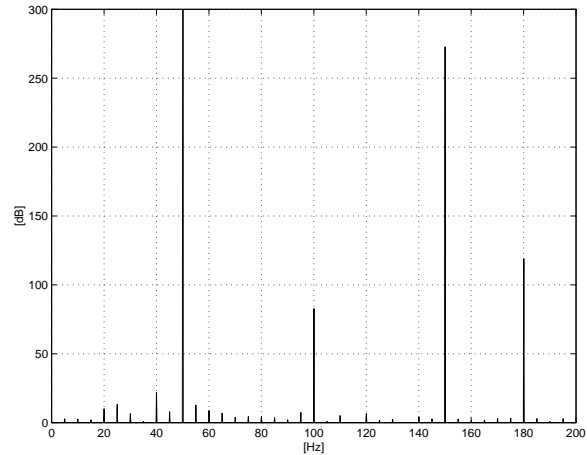,width=9truecm}
\end{center}
\caption{Normal operation frequency spectrum. The 50Hz intensity is 16667 dB}
\end{figure}
\begin{figure}[htb]
\begin{center}
\psfig{figure=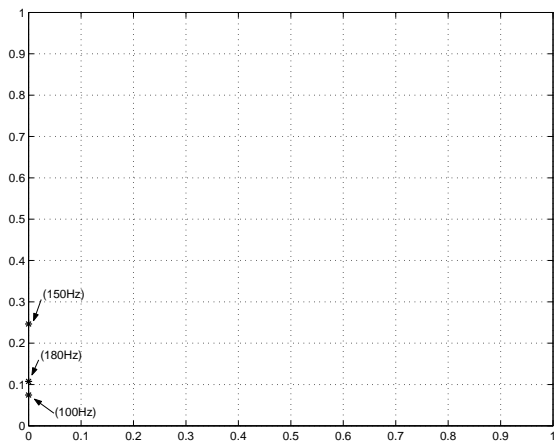,width=9truecm}
\end{center}
\caption{Three self patterns at 100, 150 and 180Hz}
\end{figure}

As shown in Fig.12, even in normal operation, the stator current spectrum
displays, in addition to the 50Hz component, other significant frequency
components (near 100, 150 and 180Hz). Asymmetry in the power supply and
misalignment are at the origin of these frequency components. Subunits
covering these frequencies detect these components as {\it self} patterns of
the motor, characterizing its normal condition state. Fig.13 shows the {\it %
self} patterns of three of these subunits. Notice that the subunits being
centered exactly at 100, 150 and 180 Hz, the slip frequency makes the harmonics
to contribute only to the integrated spectrum of the higher interval. We now
illustrate the two fault situations, namely the case of rotor broken bars
and the case of oscillatory load and unbalanced power supply.

\paragraph{Case 1: broken rotor bars}

Broken rotor bars generate spectrum lines at frequencies 
\[
f_{1}=f\pm (2ks)f 
\]
$s$ being the slip frequency, $f$ the supply frequency, and $k$ an integer
value (1, 2, 3,...). The amplitude of the sideband frequencies measures the
seriousness of the anomaly.

A subunit algorithm centered at the supply frequency ($50$Hz) has been
tested using laboratory measurements of the stator current of a
squirrel-cage motor with different numbers of broken bars.

\begin{figure}[htb]
\begin{center}
\psfig{figure=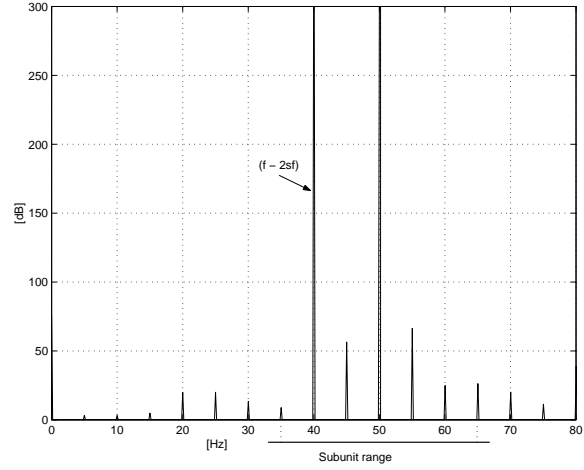,width=9truecm}
\end{center}
\caption{Motor with one broken rotor bar. Sideband frequencies and range of
the subunit around 50Hz. 50Hz intensity = 16667 dB, 40Hz intensity = 425 dB}
\end{figure}
\begin{figure}[htb]
\begin{center}
\psfig{figure=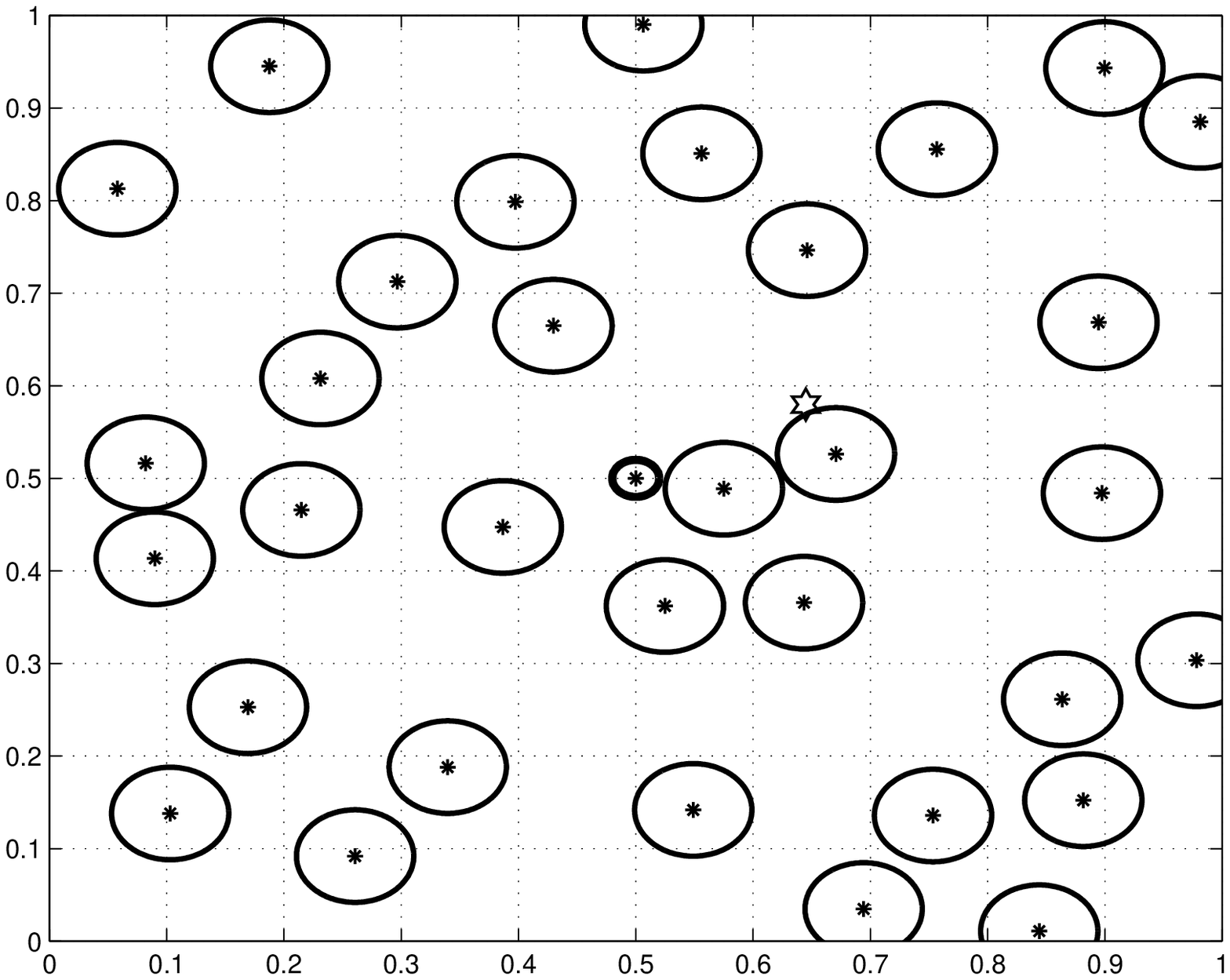,width=9truecm}
\end{center}
\caption{Initial detectors, self pattern and the external code}
\end{figure}
\begin{figure}[htb]
\begin{center}
\psfig{figure=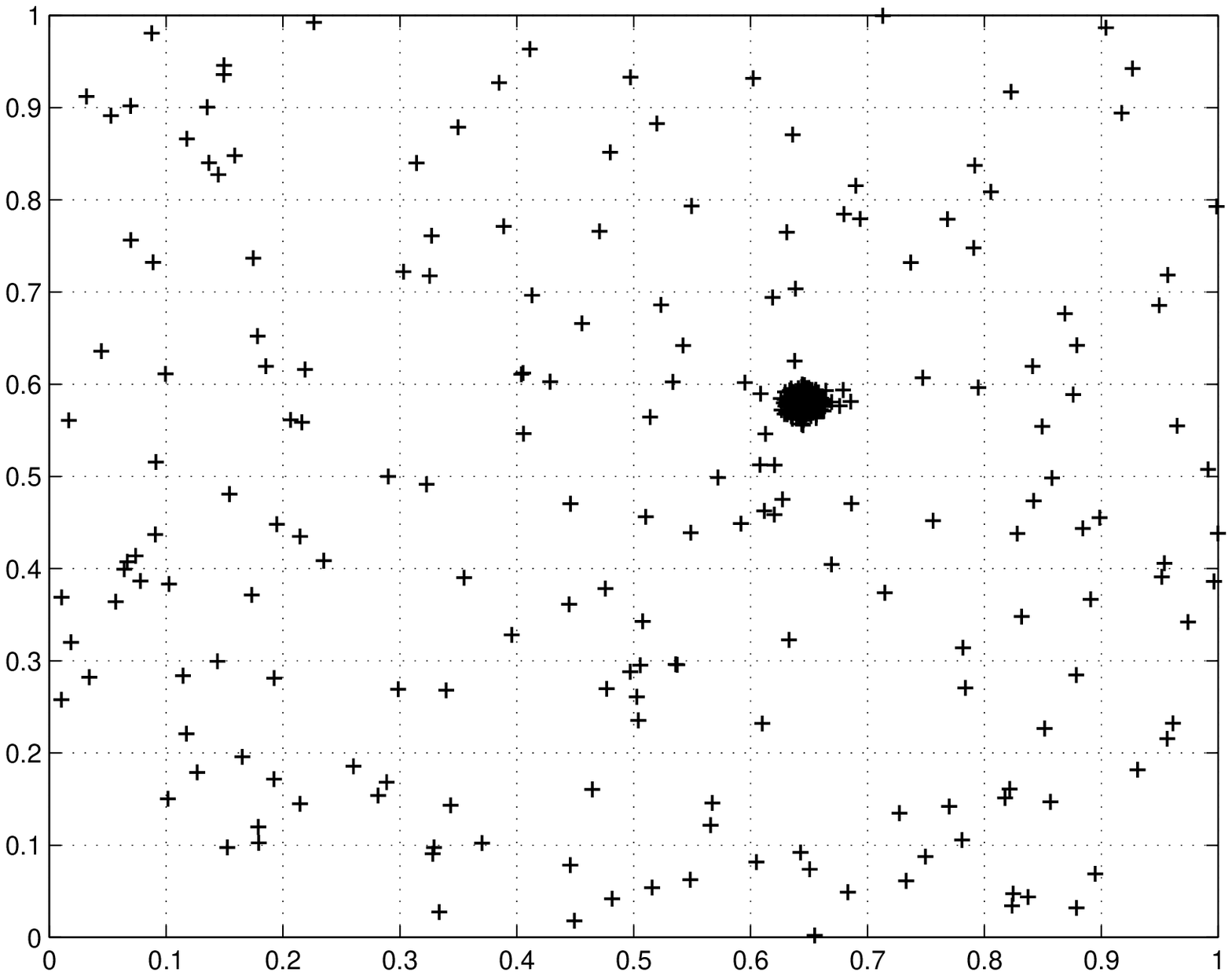,width=9truecm}
\end{center}
\caption{Vector population after the stimulation and mutation processes}
\end{figure}
\begin{figure}[htb]
\begin{center}
\psfig{figure=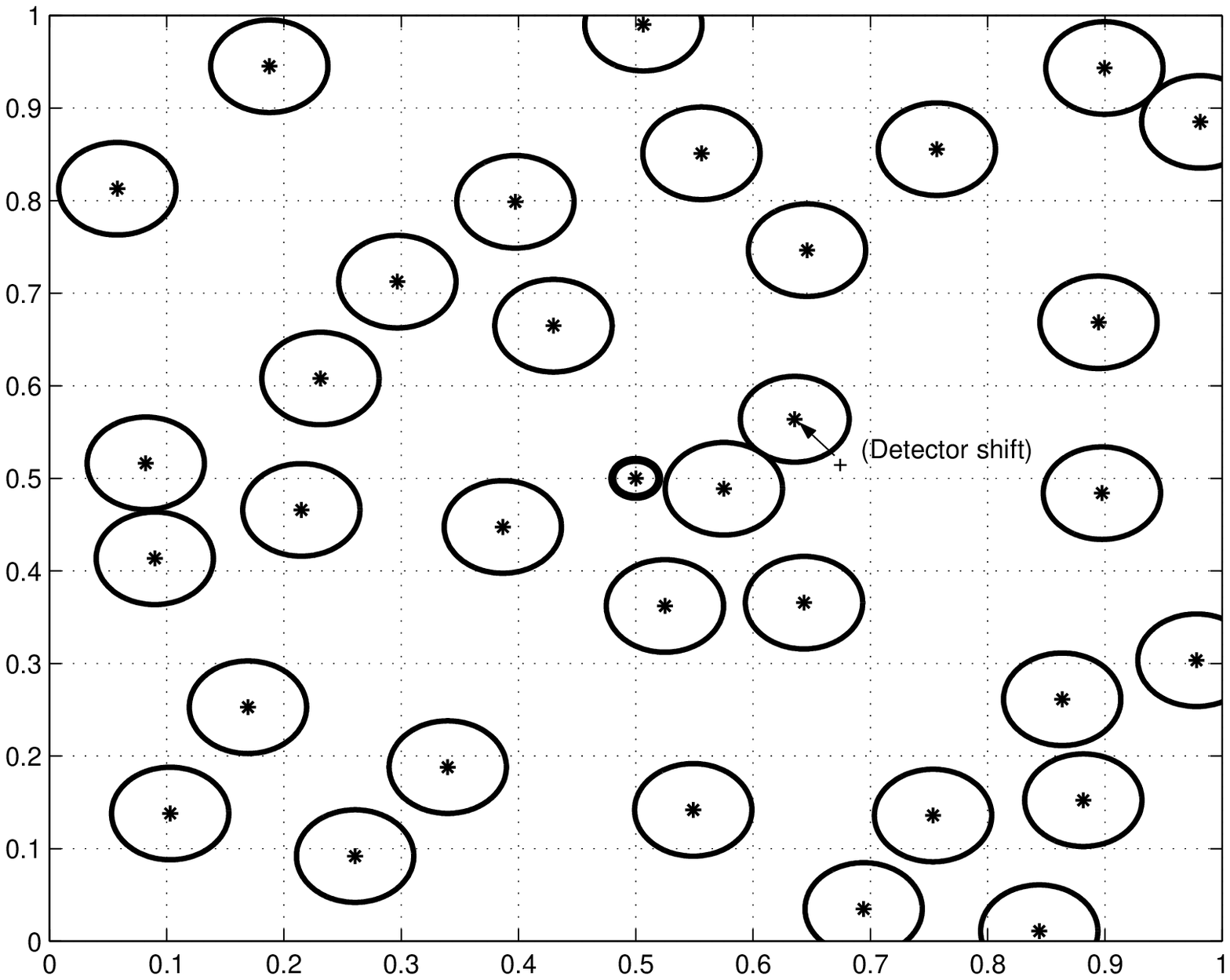,width=9truecm}
\end{center}
\caption{Detector shift}
\end{figure}

For the case of one broken bar the spectrum is shown in Fig.14. The initial
population of $T$-detectors $\left( n=33\right) $\ and the self pattern of
this subunit are shown in Fig.15. As external codes come from the system
(the hexagonal symbol in Fig.15), the $B$-module, through the stimulation
and mutation processes, creates a new distribution of the vector population
(Fig.16). The interaction between the $B$-module and the $T$-module leads to
the detector shift shown in Fig.17 and to the reporting of an anomaly.

A more severe situation is shown in Fig.18 when the motor has four broken
bars. The subunit now is the same as before. With four broken bars, the
amplitude of the sideband components has increased, creating external codes
near the boundary of state-space (Fig.18). After the stimulation and
mutation processes illustrated in Fig.19, a new detector is created (Fig.20)
to monitor and report this anomaly.

\begin{figure}[htb]
\begin{center}
\psfig{figure=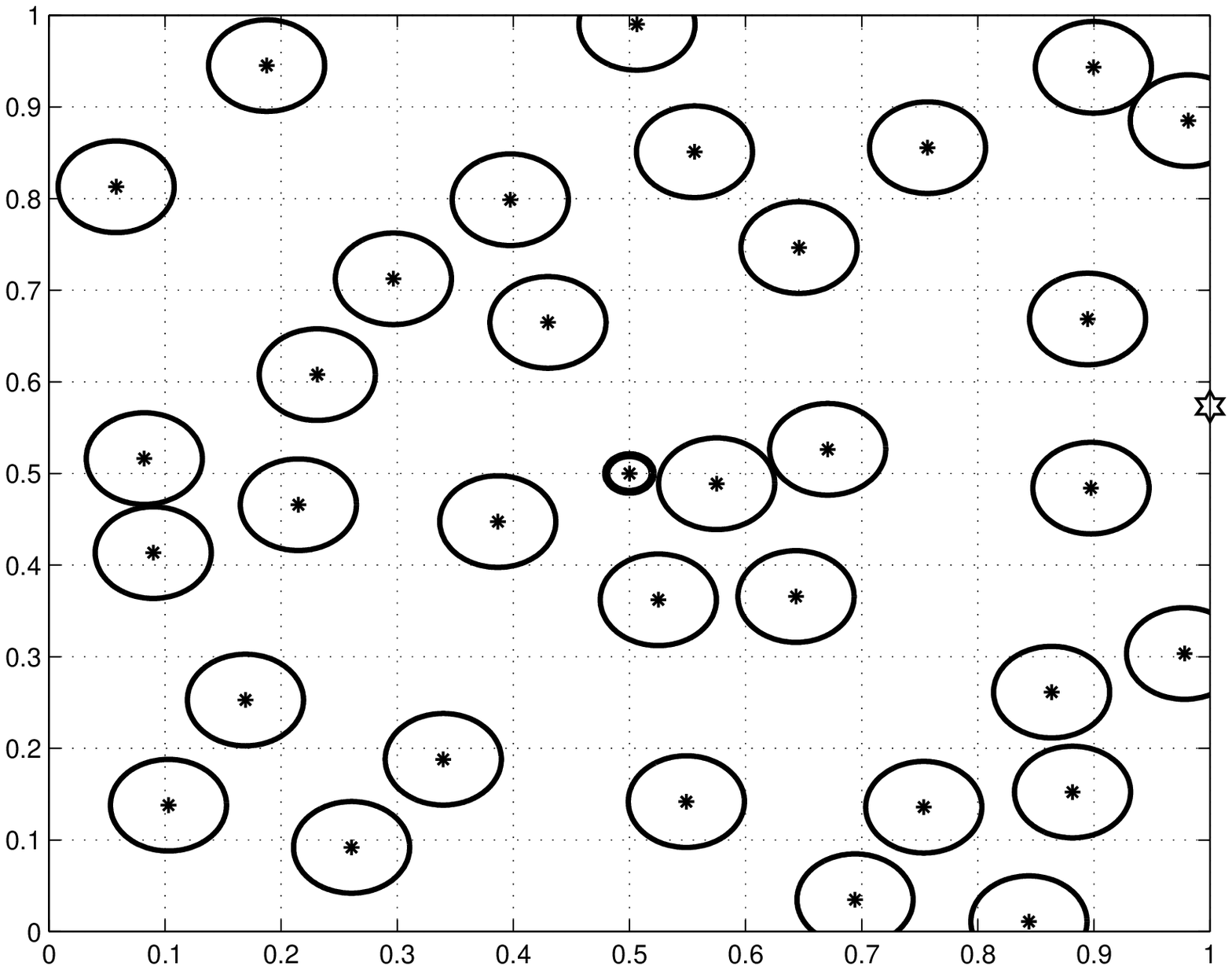,width=9truecm}
\end{center}
\caption{Motor with four broken rotor bars: Initial detectors, self pattern
and the external code}
\end{figure}
\begin{figure}[htb]
\begin{center}
\psfig{figure=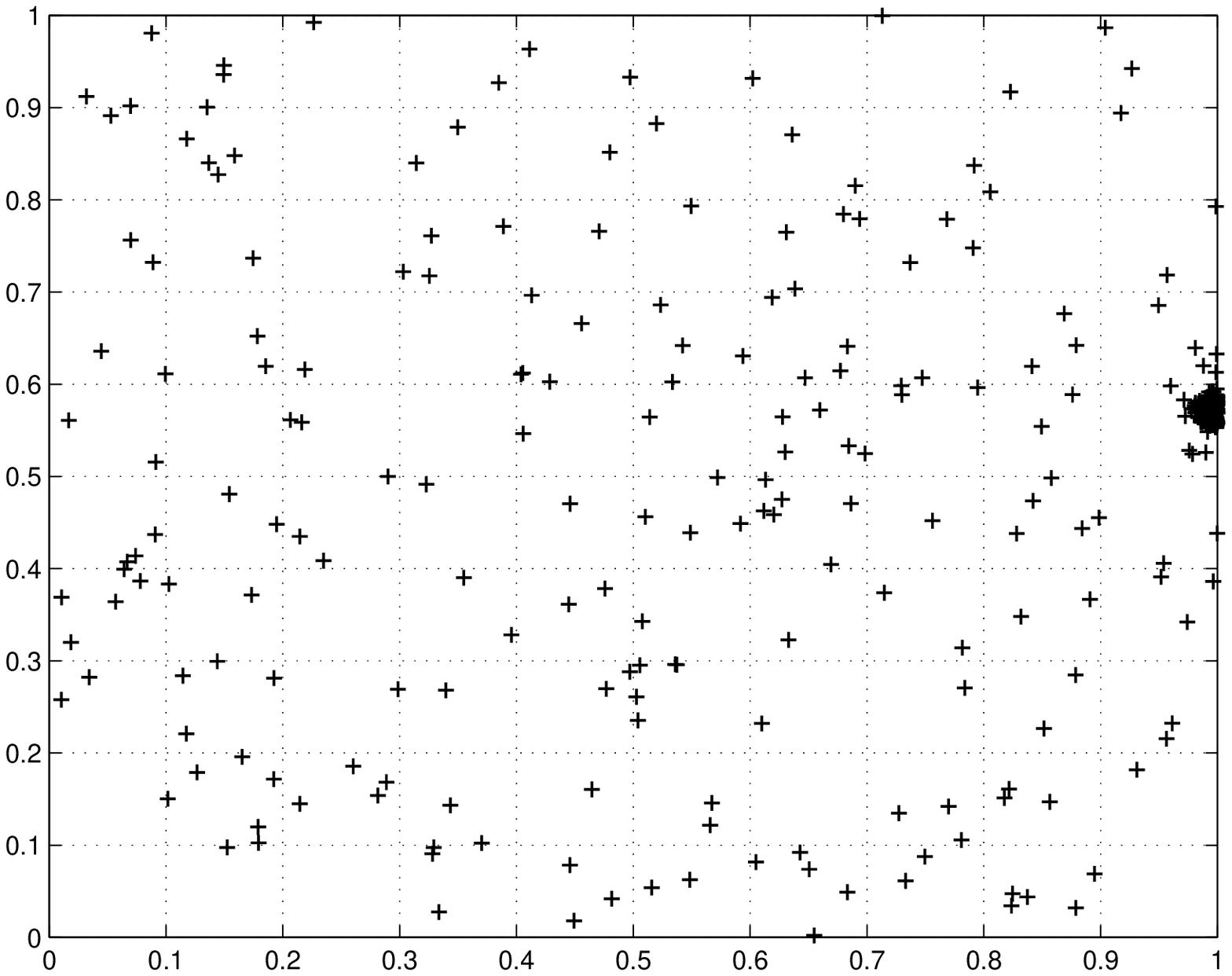,width=9truecm}
\end{center}
\caption{Vector population after the stimulation and mutation processes}
\end{figure}
\begin{figure}[htb]
\begin{center}
\psfig{figure=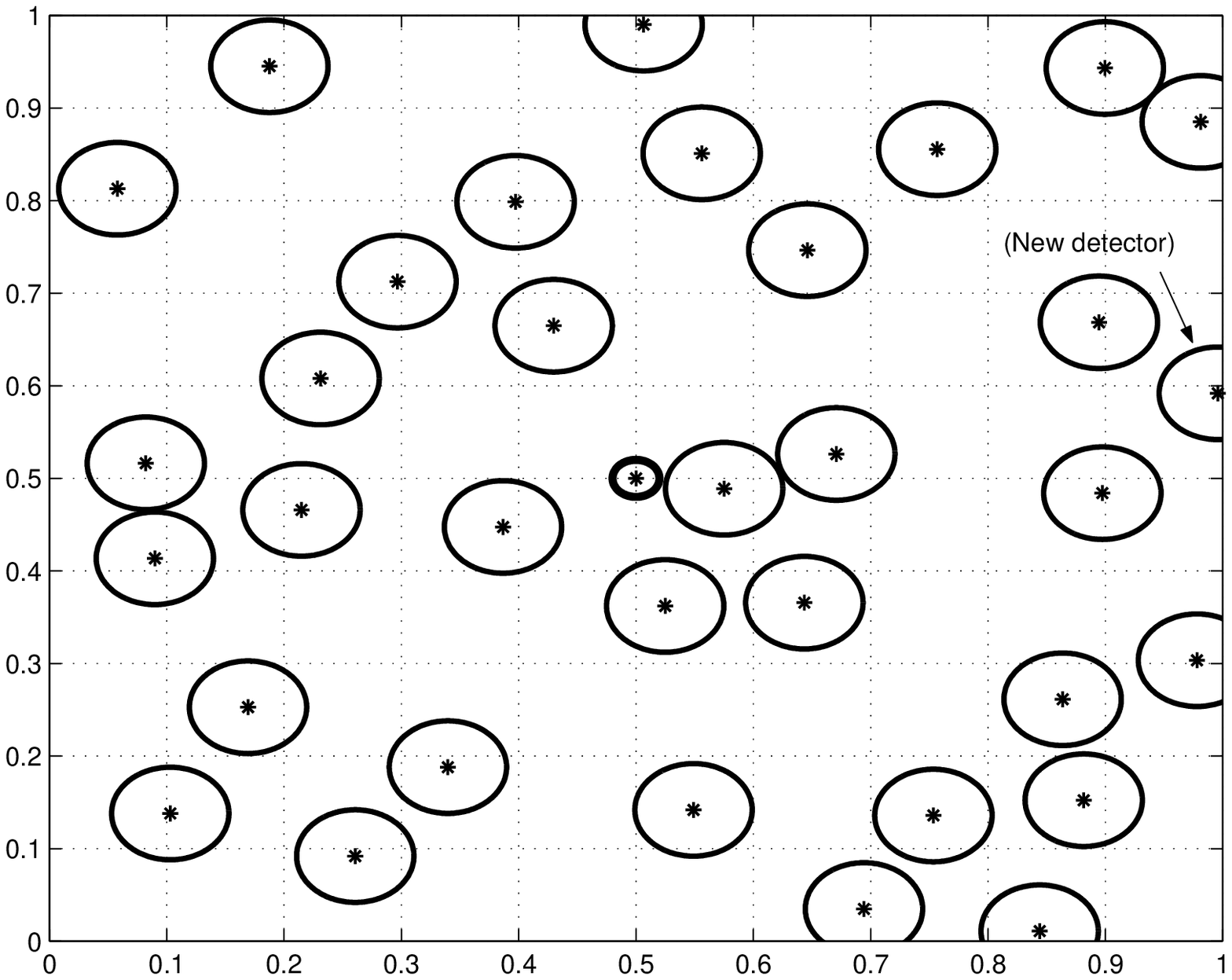,width=9truecm}
\end{center}
\caption{A new detector is created}
\end{figure}

\paragraph{Case 2: oscillatory load plus an unbalanced power supply}

An oscillation in the load torque at a multiple $m$ of the rotor speed
creates spectral lines at frequencies 
\[
f_{load}=f\left( 1\pm km\left( 1-s\right) \right) 
\]
$k=1,2,3,\cdots $. Sideband frequency components of this type are shown in
Fig.21.

\begin{figure}[htb]
\begin{center}
\psfig{figure=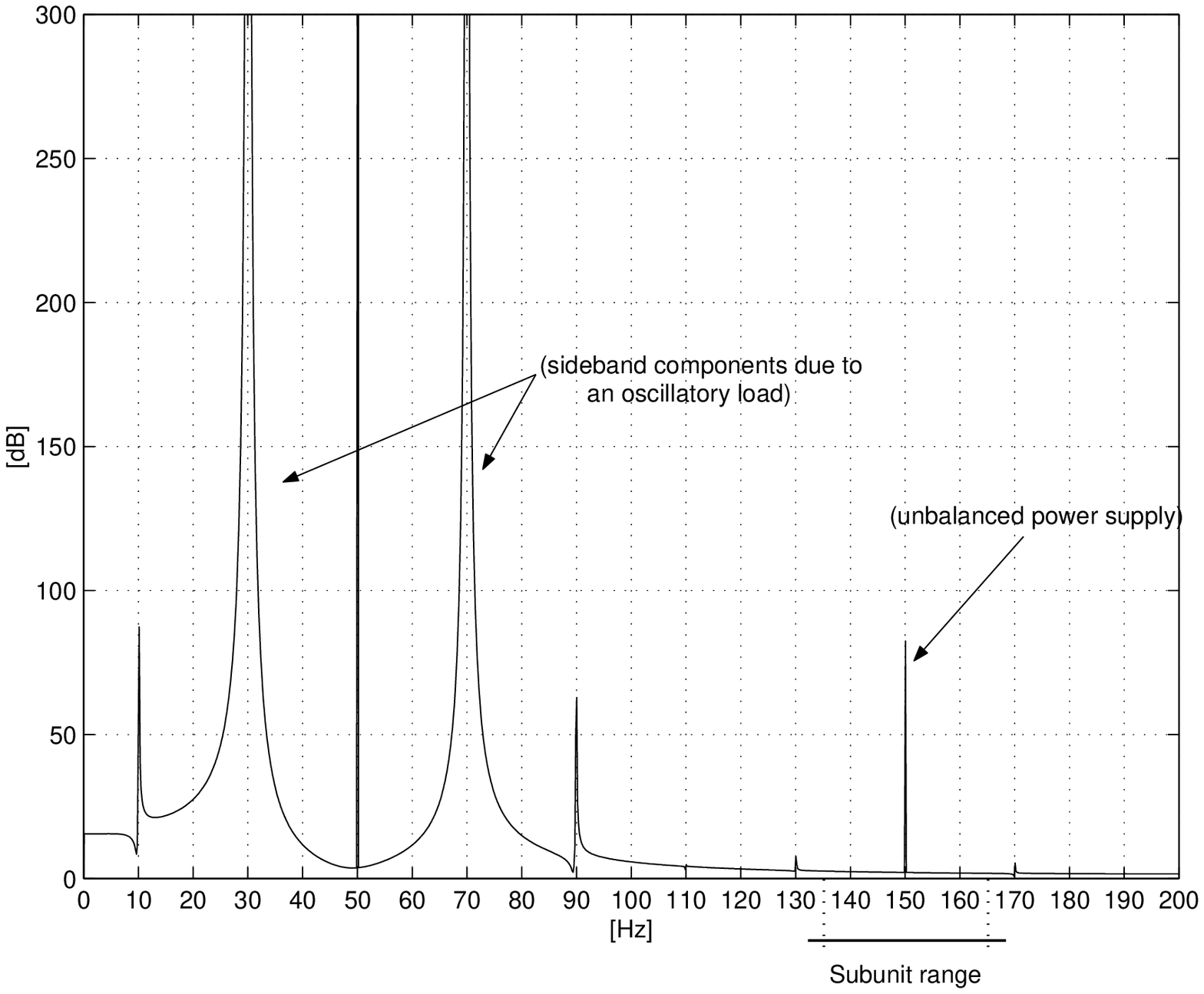,width=9truecm}
\end{center}
\caption{Motor with oscillatory load: Sideband frequency components and
range of the subunit around 150Hz. Intensities: (30Hz)=8740 dB; (50Hz)=16667 dB; 
(70Hz)=5503 dB}
\end{figure}

With an unbalanced power supply, a negative sequence voltage appears at
the stator winding. This creates frequency components 
\[
f_{umb}=f\left( 1\pm k\left( 1-s\right) \right) 
\]
A subunit centered at 150Hz detects faults of these types. Already, when
operating in normal conditions there is a small amplitude at this frequency
which characterizes the {\it self} configuration of the system. When a
situation of unbalanced operation arises, new external codes appear as shown
in Fig.22. The external codes for the unbalanced condition interact with the 
{\it B}-module resulting in the final vector population shown in Fig.23.
This then originates a new detector in the $T$-module of this subunit
(Fig.24) and an anomaly is reported.
\begin{figure}[htb]
\begin{center}
\psfig{figure=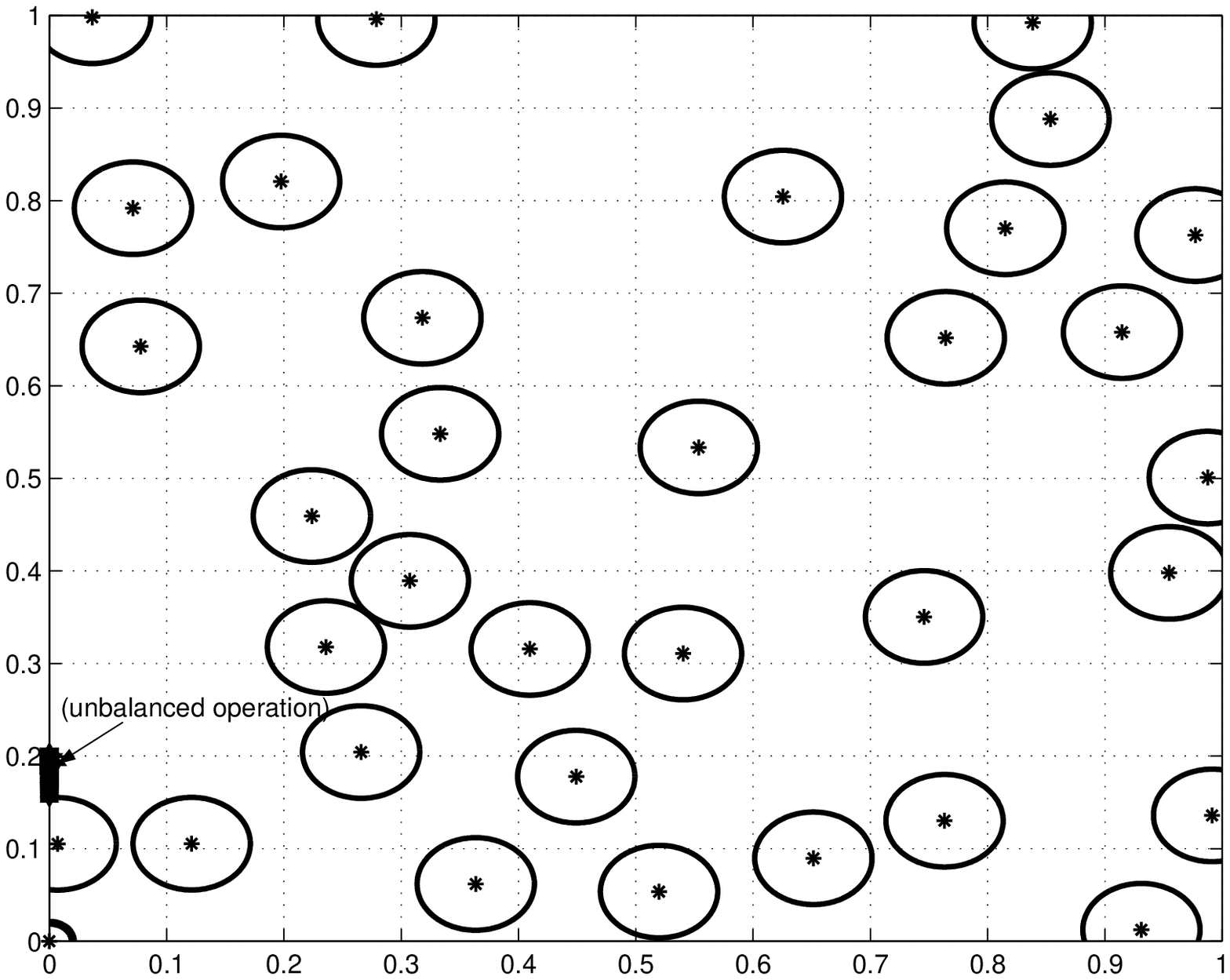,width=9truecm}
\end{center}
\caption{Initial detectors, self pattern and external codes}
\end{figure}
\begin{figure}[htb]
\begin{center}
\psfig{figure=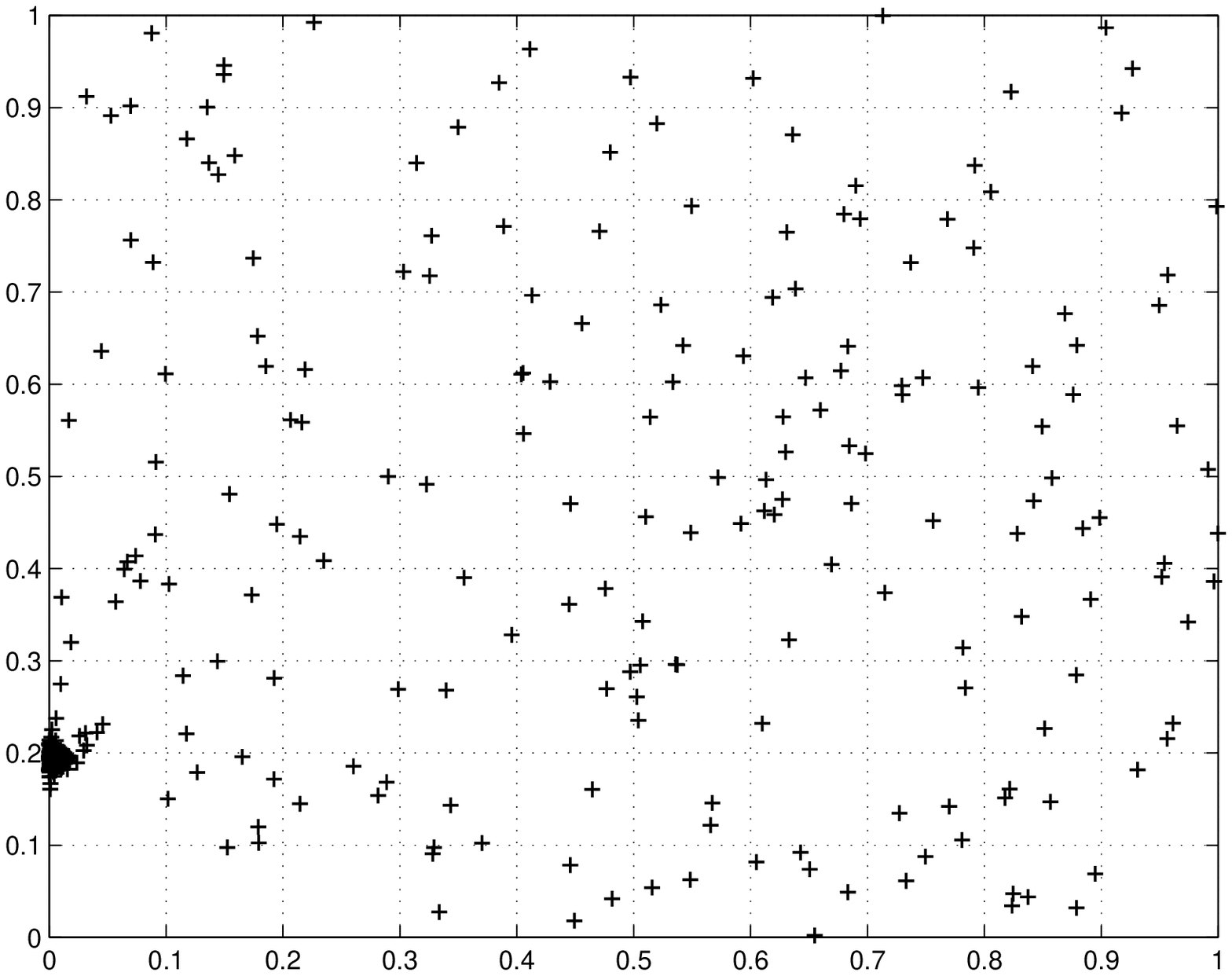,width=9truecm}
\end{center}
\caption{Vector population after the stimulation and mutation processes}
\end{figure}
\begin{figure}[htb]
\begin{center}
\psfig{figure=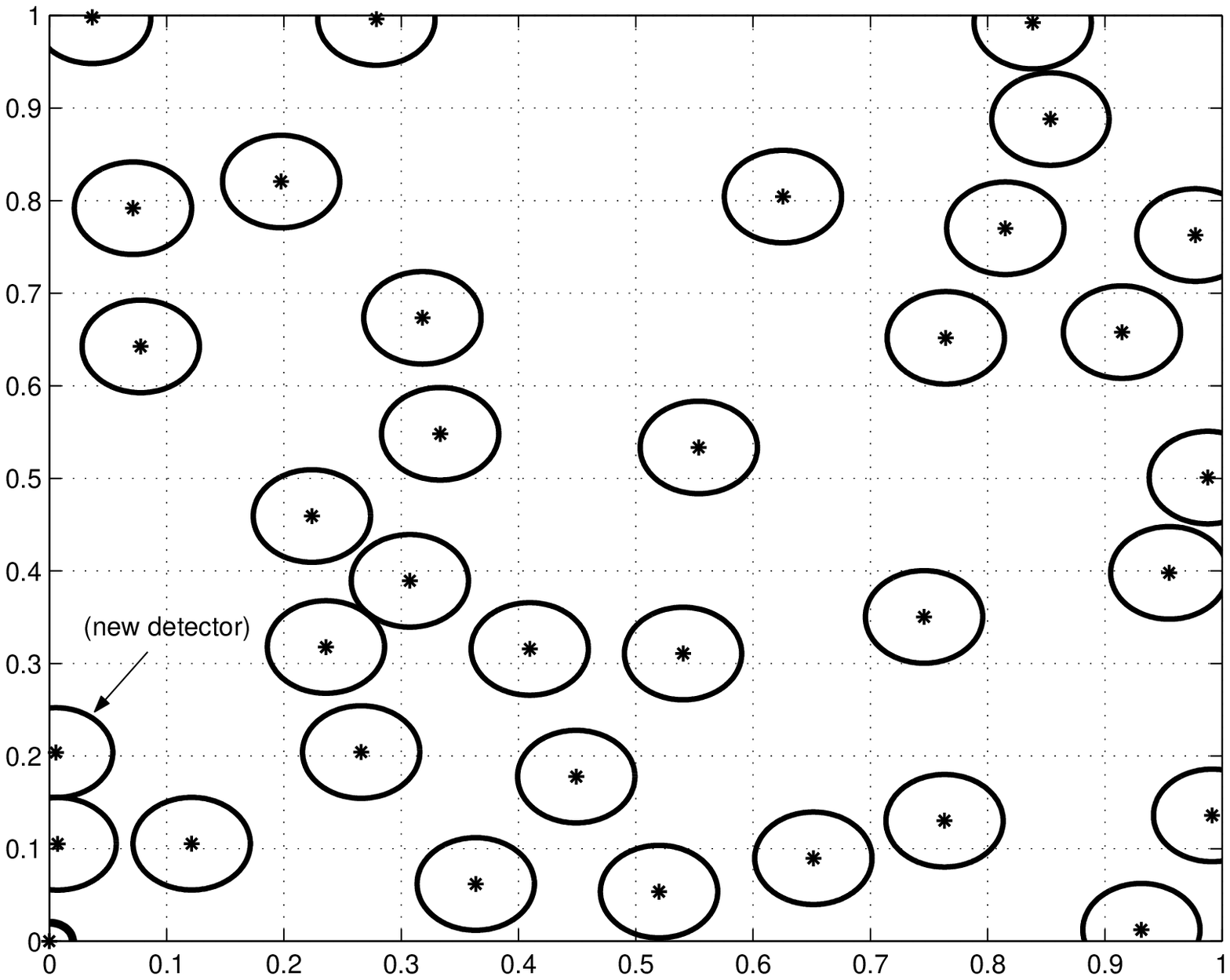,width=9truecm}
\end{center}
\caption{New detector created to monitor unbalanced operation}
\end{figure}

\section{Conclusions}

\begin{enumerate}
\item  Equipments of the same type and equal ratings have different aging
processes, depending on their particular location in the industrial process
and stress conditions. Therefore normal operating conditions of the same
type of equipment vary over a wide range of possibilities. On the other hand
it is virtually impossible to make a complete catalog of all the possible
and probable anomaly situations. By adapting itself to the actual operating
conditions of the system, fault detection based on the specific immunity
response algorithms seems to be an adequate device to characterize the
particular nature of the normal conditions as well as to react to new and
unexpected anomaly situations.

\item  By being able to detect anomaly conditions at an early developing
stage, immunity-based systems may provide substantial cost savings in
industrial processes and be an useful tool to develop preventive maintenance
schedules.

\item  In the $T$- plus $B$- module system developed in this paper, each
detected anomaly corresponds to a well defined code in the detection system.
The code of the detected anomaly is then a useful piece of information for
diagnosis and corrective measures.

\item  By trial and chance over millions of years, Nature's evolutionary
processes have found very efficient processes to deal with all kinds of
hostile environments. To obtain inspiration from these natural mechanisms
seems to be a sensible approach. However some of the features of the
biological processes are domain-specific and depend on the cell hardware
that is used. Therefore it is appropriate to obtain algorithmic inspiration
from Nature, but it would be ill advised to copy all the details of the
biological process.

\item  Organisms, as a first barrier to infection, also protect themselves
by non-specific mechanisms like macrophages, cell apoptosis, etc. These
non-specific mechanisms have a parallel with the devices used in the past
for the protection of electrical power systems (circuit breakers, partial
network shutdowns, etc.). What this analogy suggests is that it is high time
to move beyond the non-specific protection mechanisms towards specific
anomaly detection devices. Nature has been doing it for millions of years.
\end{enumerate}

{\bf Acknowledgement}

The authors are grateful to A. J. Marques Cardoso from University of Coimbra
for providing the motor fault experimental data.

\end{document}